\documentstyle[12pt,epsf,world_sci]{article}
\input math_macros
\newcommand{\MET}{\hbox{$\rlap{\kern0.25em/}E_T$} }
\newcommand{\METX}{\hbox{$\rlap{\kern0.25em/}E_{Tx}$} }
\newcommand{\METY}{\hbox{$\rlap{\kern0.25em/}E_{Ty}$} }
\newcommand{\METcal}{\hbox{$\rlap{\kern0.25em/}E_T^{cal}$} }

\newcommand{\ttbar}{$t \bar t$ }

\newcommand{\ppbar}{$p \bar p$ }

\newcommand{\emu}{$e \mu$}
\newcommand{\ee}{$ee$}
\newcommand{\mumu}{$\mu \mu$}

\def\D0{D$\!$\O }

\def\degreeC#1 {${#1}^\circ $C }
\def\Etmiss{\hbox{$\rlap{\kern0.25em/}E_T$}}
\newcommand{\etau}{$e\tau$}
\newcommand{\tautau}{$\tau\tau$}
\newcommand{\mutau}{$\mu\tau$}
\begin{document}

\title{{\bf SEARCH FOR TOP WITH \D0 DETECTOR IN DILEPTON CHANNEL}}
\author{KRZYSZTOF GENSER
\\
{\em Fermi National Accelerator Laboratory\\
Batavia, IL 60510-0500, USA\\
\vspace{0.3cm}
Representing the \D0 Collaboration}}

\maketitle
\setlength{\baselineskip}{2.6ex}

\begin{center}
\parbox{13.0cm}
{\begin{center} ABSTRACT \end{center} {\small \hspace*{0.3cm}
Preliminary results from a search for high mass \ttbar quark pair
production in $p\bar p$ collisions at $\sqrt{s} =1.8$~TeV with the \D0
detector in the \ee+jets, \emu+jets, and \mumu+jets decay channels are
presented.  No conclusive evidence for top quark production for an
integrated luminosity of~$13.5\pm1.6$~pb$^{-1}$ is observed.}}
\end{center}

\section{Introduction}

At the Fermilab Tevatron \ppbar center of mass energy the Standard
Model top quark $t$ with the mass $m_t$ greater than about
90~GeV/c$^2$ is expected to be produced mainly in \ttbar pairs through
the quark-quark annihilation with some contribution from gluon-gluon
fusion. The gluon-gluon contribution decreases as $m_t$
increases\cite{laenen}. Once $m_t$ is greater than the mass of the
{W-boson} the top quark is expected to decay via the weak charged
current ($t\rightarrow W^+b$) and the event signatures follow from the
W branching fractions.

\D0 has published a lower limit for $m_t$ of 131~GeV/c$^2$ at 95\%
confidence level\cite{d0prl1}.  The prediction for the top mass based
on the precision electroweak LEP measurements\cite{pietrzyk} is
$m_t=172^{+13+18}_{-14-20}$~GeV/c$^2$.  The CDF evidence for the top
quark indicates its mass to be\cite{cdfprl1} $m_t=174\pm
10^{+13}_{-12}$~GeV/c$^2$

The analysis presented in this paper is focused on the top mass above
120~GeV/c$^2$ and includes a new channel \mumu\ which was not used in
the previous analysis\cite{d0prl1}.  The dilepton channels considered:
\ee, \emu, and \mumu\ constitute 1/81, 2/81, and 1/81 of all \ttbar
events respectively.  The contributions from \etau, \mutau, and
\tautau\ channels having the same signatures as \ee, \emu, and \mumu\
are also taken into account in the event yields. The lepton plus jets
channels are discussed in complementary \D0
papers\cite{d0dpf2,d0dpf3}, where also the cross-section results are
summarized\cite{d0dpf3}.

\section{\D0 Detector and Data Sample}

The \D0 detector\cite{d0nim1} is based on a hermetic, high
granularity, high resolution, compensating liquid Argon -- depleted
Uranium calorimeter. The calorimeter is surrounded by a hermetic muon
system which has one super layer of proportional drift chambers before
and two super layers of chambers behind magnetized iron toroids. The
central tracking system placed between the beam pipe and the
calorimeter consists of a vertex proportional chamber, transition
radiation detector and central, and forward drift chambers. There is no
central magnetic field.  The calorimeter energy resolutions can be
parametrized as $\sigma(E) /E
\approx {\it A} / \sqrt{E}$ ($E$ in GeV), where {\it A}=0.15 for
electrons, and {\it A}=0.80 for jets.  The muon momentum resolution is
$\sigma\left({1/p}\right)\approx {0.2/p} + 0.01$~GeV/c$^{-1}$.  For
minimum bias events, the resolution for each component of the missing
transverse energy \MET is about 1.1~GeV+0.02$\times$($\Sigma E_T$),
where $\Sigma E_T$ is the scalar sum of all the transverse energy
$E_T$ in the calorimeter.

The data sample analyzed was collected during the Fermilab Tevatron
Collider run 1992/93 and has a corresponding luminosity
of~$13.5\pm1.6$~pb$^{-1}$ for the \ee\ and \emu\ channels, and
$9.8\pm1.2$~pb$^{-1}$ for the \mumu\ channel.

\section{Lepton Identification}

All the \ttbar\ dilepton decay channels are characterized by the
presence of two high transverse momentum isolated leptons, large
missing transverse momentum and significant jet activity. Once
appropriate lepton identification and kinematic cuts are applied one
can expect a good signal to background ratio and a good efficiency.

\subsection{Electron Identification}

Electrons are identified as energy clusters in the electromagnetic
calorimeter within the pseudorapidity region of $|\eta| < 2.5$. The
clusters are required to have at least 90\% of their total energy in
the electromagnetic part of the calorimeter. The longitudinal and
transverse cluster profile has to be consistent with the shape of the
electromagnetic shower initiated by an electron\cite{hmatrix}. The
isolation requirement is \unskip\break
$(E_{TOT}^{0.4}-E_{EM}^{0.2})/E_{EM}^{0.2}<0.1$ where $E_{TOT}^{0.4}$
and $E_{EM}^{0.2}$ are the total and electromagnetic cluster energy
contained within the cone of ${\cal R} < 0.4$ and ${\cal R} < 0.2$
respectively and ${\cal R}$ is defined as ${\cal R} =
\sqrt{(\Delta\phi)^2 + (\Delta\eta)^2 }$.
It is also required that there is a central detector track coming from
the interaction vertex which is matched with each cluster. In the case
of the \ee\ channel, the track has to have the corresponding energy
deposition in the drift chambers which is consistent with that of a
single charged particle.

\subsection{Muon Identification}

Muons are identified as tracks in the muon drift chambers after
penetrating through 13 to 19 interaction lengths of calorimeter and
muon toroids.  It is required that there is an associated minimum
ionizing deposition of at least 0.5~GeV in the calorimeter along the
path of the muon if there is a central track matching the muon track,
and a deposition of at least 1.5~GeV if there is not a central
detector track which matches the muon track.  To reject cosmic rays,
no track or hits consistent with a track back-to-back in $\eta$ and
$\phi$ to the considered muon are allowed.  It is required that the
magnetic field integrated over the path of the muon is greater than
1.83~Tm in order to assure good momentum determination.  In case of the \emu\
channel, the isolation is imposed by rejecting muons for which there is
a jet with the transverse energy $E_T$ greater than 8~GeV within the
cone of ${\cal R} < 0.5$ or for which there is an energy deposit of
4~GeV or more in an annular cone of $0.2<{\cal R} <0.4$ around the
muon direction. For the \mumu\ channel, to impose isolation, muons
with the transverse momentum relative to the nearest jet smaller than
5~GeV/c are rejected.  For the \emu\ channel, all muons within
pseudorapidity region of $|\eta| < 1.7$ are considered. For the \mumu\
channel, central muons ($|\eta| < 1.1$) are used.

\section{Analysis of Dilepton Channels and Background Processes}

\subsection{\mumu\ channel}

This is a new channel which was not included in the previous
analysis\cite{d0prl1}.  The trigger for this channel requires a muon
candidate with the transverse momentum $p_T$ greater than 14~GeV/c and
a jet candidate with $E_T>15$~GeV.  The offline cuts demand two muons,
each with $p_T>15$~GeV/c. To reject background processes, two jets
with $E_T>15$~GeV are required. Dimuon invariant mass $M_{\mu\mu}$ is
required to be greater than 10~GeV/c$^2$ to remove $J/\Psi$ or $\Psi'$
events. $Z^0\rightarrow\mu\mu$ events are rejected by demanding that
the angle between \MET and the leading muon to be smaller than
$165^{\circ}$ ($165^{\circ}$) for two (three) layer muon tracks. Since
the calorimeter missing transverse energy
\METcal is an independent measure of the transverse momentum of the muon
pair, the events are rejected if $\Delta \phi (\METcal,\vec{p}_{T}^{\mu
\mu} ) > 30^{\circ}$ where $\Delta\phi$ is the azimuthal opening angle.
Further rejection of the $Z^0\rightarrow\mu\mu$ events is achieved by
the cut \MET $>$ 40 GeV when
$\Delta\phi(\vec{p}_{T}^{\mu_1},\vec{p}_{T}^{\mu_2}) > 140^{\circ}$.
No events remain after all the cuts. Very good agreement between the
data and the dominant background process $Z^0\rightarrow\mu\mu$ is
seen at all the stages of the analysis. The predicted number of
$Z^0\rightarrow\mu\mu$ events is $0.28\pm0.05$. The event yields for
this and for the other channels, for the data, \ttbar Monte Carlo and
for the background processes are summarized in table~1. The event
yields for the \ttbar production were calculated using central value
of the \ttbar cross-section\cite{laenen}. The number of background
events coming from processes of the same signature as the process
under study was predicted using Monte Carlo simulations. The errors
quoted are statistical and systematic. In addition to the quoted
errors there is an additional normalization error due to the 12\%
uncertainty in the value of the integrated luminosity.

\subsection{\ee\ channel}

The trigger used in this channel is a logical OR of triggers demanding
one electron candidate with $E_T>20$~GeV, or two electron candidates
with $E_T>10$~GeV, or an electron candidate with $E_T>20$~GeV and
\MET$>20$~GeV, or one electron candidate with $E_T>15$~GeV and
\MET$>20$~GeV and two jets with $E_T>16$~GeV. The offline
selection cuts require two electrons, each with $E_T>20$~GeV,
\MET$>25$~GeV and two jets with $E_T>15$~GeV. $Z^0\rightarrow\ ee$
events are rejected by demanding $|M_Z-M_{ee}|>12$~GeV/c$^2$ if
\MET$<40$~GeV. No events remain after all the cuts.
The complete event yields can be found in table~1.  After all cuts, the
remaining dominant background processes are $Z^0\rightarrow\ \tau\tau
\rightarrow\ ee$ and $Z^0\rightarrow\ ee$. There is also an
instrumental background when a jet is misidentified as an electron in
the W+jets events. The probability of a jet faking an electron was
estimated using an independent jet data sample.

\subsection{\emu\ channel}

The trigger used in this channel is also a logical OR of triggers
requiring one electron candidate with $E_T>7$~GeV and one muon
candidate with $p_T>5$~GeV/c, or a muon candidate with $p_T>14$~GeV/c
and a jet candidate with $E_T>15$~GeV, or an electron candidate with
$E_T>15$~GeV and \MET$>20$~GeV and two jets with $E_T>16$~GeV. The
offline kinematic cuts require muon $p_T>12$~GeV/c, electron
$E_T>15$~GeV, {\MET$>10$~GeV}, and two jets with $E_T>15$~GeV.  To
reject W($\mu\nu_{\mu}$)+jets events it is required that
\METcal$>20$~GeV.  The $\Delta R_{e\mu}>0.25$ cut is used to remove muon
bremsstrahlung events.  One event survives all the cuts.  This is the
same remarkable event which was found in the previous analysis.  The
event is characterized by two very high $p_T$ leptons. Figure~1a shows
muon $1/p_T$ versus electron $E_T$ for the data before the two jet cut
is applied. The surviving event is marked by a $\star$. Figure~1b
shows the corresponding Monte Carlo distribution. The event yields are
summarized in table~1. The remaining dominant background processes are
$Z^0\rightarrow\ \tau\tau \rightarrow\ e\mu$ and
W($\mu\nu_{\mu}$)+jets with one jet misidentified as an electron.

\vskip-2.0in
\hbox{\hskip-.5in
\epsfysize=8.5in
\epsfbox{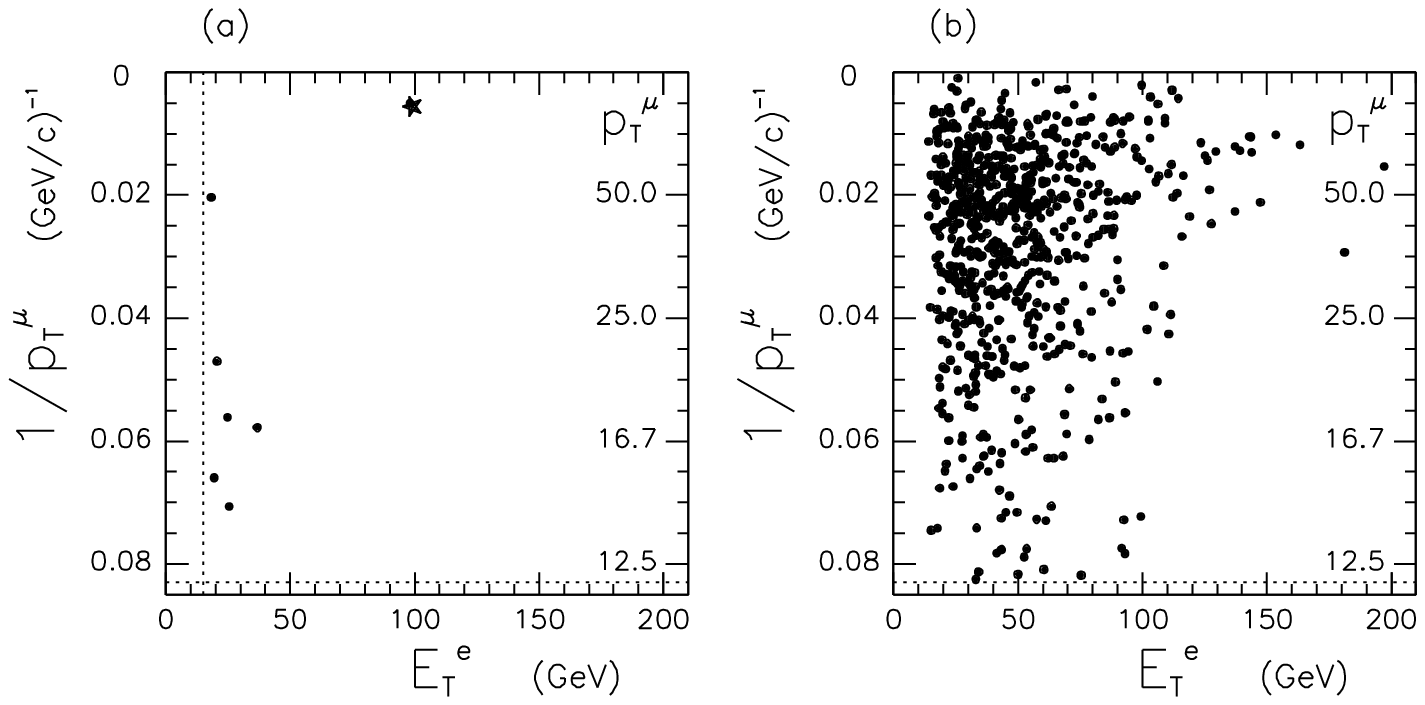}}
\vskip-4in
{\small Fig.\ 1.\  Muon $1/p_T$ versus electron $E_T$ for the \emu\
channel; a) Data 13.5~pb$^{-1}$, b) 170~GeV/c$^2$ \ttbar Monte Carlo
21.3~fb$^{-1}$.}
\vfill
\begin{center}
{\small Table 1.  Summary of event yields. The errors quoted are
statistical and systematic. In addition to the quoted errors there is
an additional normalization error due to the 12\% uncertainty in the
value of the integrated luminosity.}
\end{center}
\begin{center}
\begin{tabular}[p]{|c|c|c|c|}
\hline
 & & & \\
\multicolumn{1}{|c}{ } & \multicolumn{1}{|c|}{ \emu }&
\multicolumn{1}{c|}{ \ee }& \multicolumn{1}{c|}{ \mumu } \\
 & & & \\
 Luminosity $[$pb$^{-1}]$ & 13.5 $\pm$ 1.6 & 13.5 $\pm$ 1.6 & 9.8 $\pm$ 1.2
 \\ \hline
 $t\bar{t}$ MC & & & \\
 140 GeV/c$^2$ &
0.72 $\pm$ 0.16 &
0.41 $\pm$ 0.07 &
0.24 $\pm$ 0.05 \\
 160 GeV/c$^2$ &
0.40 $\pm$ 0.09 &
0.22 $\pm$ 0.04 &
0.12 $\pm$ 0.02 \\
 180 GeV/c$^2$ &
0.23 $\pm$ 0.05 &
0.12 $\pm$ 0.02 &
0.06 $\pm$ 0.01 \\
\hline
 Data &
1  & 0 & 0 \\
\hline
 Background &
0.27 $\pm$ 0.09 &
0.16 $\pm$ 0.07 &
0.33 $\pm$ 0.06 \\ \hline
\end{tabular}
\end{center}
\vfill
\eject

\section{Conclusions}

\D0 has performed a search for high mass $t\bar{t}$ decays into the
three dilepton decay modes: \emu+jets, \ee+jets, and \mumu+jets. One
event survives in the $e\mu$ channel.  This is the same event as found
in the original \D0 analysis\cite{d0prl1}.  No candidates are found in
the \ee\ and \mumu\ channels. Since the number of expected background
events is consistent with the one event seen, it is concluded that no
significant \ttbar signal is observed. The results are preliminary.

\bibliographystyle{unsrt}

\end{document}